\documentclass{elsart}
\usepackage{harvard}
\usepackage{amssymb,epsf}
\usepackage{graphicx}

\makeatletter
\def\elsartstyle{%
	\def\normalsize{\@setfontsize\normalsize\@xiipt{14.5}}
	\def\small{\@setfontsize\small\@xipt{13.6}}
	\let\footnotesize=\small
	\def\large{\@setfontsize\large\@xivpt{18}}
	\def\Large{\@setfontsize\Large\@xviipt{22}}
	\skip\@mpfootins = 18\p@ \@plus 2\p@
	\normalsize
}
\makeatother

\def\url#1{{\ttfamily\def\/{/\discretionary{}{}{}}#1}}

\newcommand{\be}{\begin{equation}}
\newcommand{\ee}{\end{equation}}
\def\bea{\begin{eqnarray}}
\def\eea{\end{eqnarray}}

\newcommand{\tc}{{$t_{solve}$~}}  
\newcommand{\tr}{{$t_{tree}$~}}  
\newcommand{\tp}{{$t_{solve}(P)$~}}

\newcommand{\La}{{$\Lambda$CDM}}
\newcommand{\gr}{\kern 2pt\hbox{}^\circ{\kern -2pt K}} 
\newcommand{\oml}{\Omega_{\Lambda}}
\newcommand{\omm}{\Omega_{m}}
\newcommand{\omb}{\Omega_{b}}
\newcommand{\brr}{\begin{array}}
\newcommand{\err}{\end{array}}
\newcommand{\ltsima}{$\; \buildrel < \over \sim \;$}
\newcommand{\simlt}{\lower.5ex\hbox{\ltsima}}
\newcommand{\gtsima}{$\; \buildrel > \over \sim \;$}
\newcommand{\simgt}{\lower.5ex\hbox{\gtsima}}

\begin{document}
\begin{frontmatter}

\title{Parallelization of a treecode }

\author{{\thanksref {valda}} Riccardo Valdarnini}
\thanks[valda]{E--mail: valda@sissa.it}

\address{SISSA -- International School for Advanced Studies,
    Via Beirut 2/4, Trieste, Italy}

\begin{abstract}
I describe here the performance of a parallel treecode with individual 
particle timesteps. The code is  based on the Barnes-Hut algorithm and runs 
cosmological N-body simulations on parallel machines with a distributed memory 
architecture using the MPI message-passing library. For a configuration with a 
constant number of particles per processor the scalability of the code  was
tested up to $P=128$ processors on an IBM SP4 machine. In the large $P$ 
limit 
the average CPU 
time per processor necessary for solving the gravitational interactions is 
$\sim 10 \%$ higher than  that expected from the ideal scaling relation.
The processor domains
are determined every large timestep according to a recursive orthogonal 
bisection, using  a weighting scheme 
which takes into account the total particle computational load within the 
timestep.
The results of the numerical tests show that the load balancing efficiency 
$L$ of the code is high ($\simgt90\%$)  up to $P=32$, and decreases to 
$L\sim 80\%$ when $P=128$. In the latter case it is found that some aspects
of the code performance are affected by machine hardware, while the
proposed weighting scheme can achieve a load balance as high as $L\sim 90\%$ 
even in the large $P$ limit.

\end{abstract}
\end{frontmatter}
%

\section{Introduction}
\label{sec:introduction}

According to the standard picture the observed structure in the universe has
arisen via gravitational instability from the time evolution of initial 
irregularities in the matter density distribution. These perturbations grew 
under their own gravity and during early epochs their cosmic evolution can be 
described according to standard linear perturbation theory. At late 
epochs the evolution of structure on scales of relevant cosmological 
interest is characterized 
by nonlinearity. 
Numerical simulations play a fundamental role in improving the theoretical 
understanding of structure formation. 
This approach 
has received a large impulse from the huge growth of computer technology 
in the last two decades. Cosmological N-body simulations are now widely
used as a fundamental tool in modern cosmology for testing viable theories 
of  structure formation. The most important task of cosmological codes is the 
computation of the gravitational forces of the system. Several methods have 
been developed to solve the large-scale gravitational field (see, e.g., 
Bertschinger 1998 for a review). A popular approach is the tree algorithm
\cite{ap85,he87}. 
The particle distribution of the system is arranged into 
a hierarchy of cubes and the force on an individual particle is computed 
by a summation of the multipole expansion of the cubes. The main advantage 
of a tree-based code is the computational cost of a force evaluation, 
which scales with the particle number $N_p$ as $\propto N_p \log N_p$. 
 This makes tree 
codes particularly indicated in simulations where 
a large number of particles is required. 
Another important point in favour of tree codes is that individual 
timesteps for all of the particles can be implemented easily, which allows a 
substantial speed-up  of the force evaluation for a clustered distribution. 
For a typical evolved distribution the particles located in high-density 
regions will need to be advanced using the smallest timesteps, but they 
represent only a small fraction of the total particle number.
 If the particles are advanced with a single timestep $\Delta t$, the overall 
accuracy in the orbit integration is then maintained only if $\Delta t$ is the 
minimum timestep $\Delta t_{min}$  of the particle configuration.  
In the single-step integration scheme the force of all the particles 
is re-evaluated at each step, but if $\Delta t_{min}$ becomes very small this 
implies that a large amount of computational work is wasted in calculating 
the force of particles in regions where the imposed global accuracy is not 
required.
 If individual timesteps are allowed, the accelerations are then evaluated 
at each step only for those particles which have been identified according 
to a specified stability criterion.

 An important task is the improvement of the dynamic range of the 
simulations. Large simulation volumes are required for statistical 
purposes but, at the same time, modeling the formation and evolution 
of each individual galaxy in the simulated volume requires that a 
realistic simulation should be implemented with $10^8 \sim 10^9$ 
particles.
This computational task can be efficiently solved if the code is adapted 
to work on a parallel machine where many processors are linked together 
with a communication network.  
If the architecture of the machine is memory-distributed the optimal code
should distribute the computational load in an even way on all the 
processors and, at the same time, minimize the communications among all the 
processors. Because of the long-range nature of gravity the latter task is 
inherently difficult. On the other hand, with respect to a serial code for
a network of $P$ processors the computation reward in terms of CPU time 
is expected to be close to $\sim P$. These arguments have led a number of 
authors to parallelize treecodes \cite{sa91,wa94,du96,da97,li00,sp01,mi02}.
In this paper I present a parallel implementation of a multistep 
treecode based on the Barnes-Hut (1986, BH) algorithm. 
 The code is cosmological and uses the MPI message library.
The paper is organized as follows.
In sect. 2 the tree algorithm is presented and the dependence of the 
acceleration errors on the tree-opening parameter $\theta$ is discussed,
together with the implemented scheme for the parallelization of the treecode. 
 Parallel performances are discussed in sect. 3 and the main
results are summarized in sect. 4.

\section{Parallelization of a treecode}
The BH algorithm works by subdividing a root box of size $L$, which contains 
all of the simulation particles, into 8 subvolumes of size $L/2$. This procedure
is then repeated for each of the subcubes and continues until the remaining 
cells or nodes
 are empty or have one particle. After the $k-th$ iteration the size 
of the subcubes is $l_k=L/2^k$. 
After the tree construction is complete
the multipole moments of the mass distribution inside the cells are 
computed starting from the smallest cells and  
 proceeding up to the root cell. 
The moments of the cells are typically approximated up to quadrupole order.
For each particle the acceleration is  
evaluated by summing the contribution of all of the cells  and particles
which are in an interaction list. The list is constructed starting
from the root cell and descending the tree down to a required 
level of accuracy. At each level a cell of the tree is accepted 
if it satisfies an accuracy criterion. 
If the cell fails this criterion then it is opened, the  
particles contained are added to the interaction list and the accuracy 
criterion 
is applied again for the remaining subcells.
BH have introduced the following acceptance criterion 
for the examined cell
\begin{equation}
d> l_k/\theta,
\label{op}
\end{equation}
where $d$ is the distance between the center of mass (c.o.m.) of the cell and 
the particle position, $\theta$ is an input parameter which controls the 
accuracy of the force evaluation. 
When $\theta$ gets smaller more cells are 
added to the interaction lists and this implies a more accurate force 
evaluation.
For $\theta \rightarrow 0$ one recovers the direct summation.
When $\theta$ is large ($\simeq 1$) and the c.o.m. is close to the edge of the 
cell 
an acceptance criterion which avoids possible errors in the force calculation 
is given by \cite{ba94,du96}
\be
d> l_k/\theta+\delta,
\label{opn}
\ee
where $\delta$ is the distance between the cell c.o.m. and its geometrical 
center. 

\subsection{Domain decomposition}
 The spatial domains of the processors are determined according to the 
orthogonal recursion bisection (ORB, Salmon 1991). The computational volume is 
first cut along the x-axis at a position $x_c$ such that 
\be
\sum_{i<} w_i \simeq \sum_{i>} w_i,
\label{orb}
\ee
 where the summations are over all of the 
particles 
with $x_i <x_c$ or $x_i > x_c$ and $w_i\propto N_{OP}(i)$
 is a weight assigned to 
each particle proportional to the number of floating point operations
 (i.e. the computational work) which are necessary to compute the particle 
force.

When the root $x_c$ has been determined the particles are then exchanged 
between the processors until all of the particles with $x_i <x_c$ belong 
to the first $P/2$ processors, and those with $x_i > x_c$ are in the second
 $P/2$ processors.
The whole procedure is repeated recursively, cycling through the cartesian
dimensions, until the total number of subdivisions of the computational 
volume is $log_2 P$ ( with this algorithm $P$ is constrained 
to be a power of two).
At the end of the domain decomposition, the subvolumes will enclose a 
subset of particles with approximately an equal amount of computational 
work. The calculation of the forces is then approximately 
load-balanced among all of the processors.

The parallel treecode presented here uses individual particle timesteps 
and the number of active particles $N_{act}$ for which it is necessary
 to calculate the forces is
highly variable with the current timestep. This number at a given time can be 
a small fraction of the total particle number. 
In such cases the load balancing scheme, which has been determined 
according to the ORB algorithm at the previous step, can be significantly 
degraded. One solution is to perform the ORB at each timestep, but this
implies a communication overhead when $N_{act}$ is small.
In the multistep integration scheme $\Delta t_0$ is the maximum allowed 
particle timestep and the particle positions are synchronized when the
simulation time $t$ is a multiple of $\Delta t_0$; see, e.g., Hernquist
\& Katz (1989) for more details.
In the parallel treecode presented here, the ORB domain decomposition 
is applied  every large timestep $\Delta t_0$, when the simulation time $t$ is 
$n \Delta t_0$. 
The load-balancing of the code is presented in sect. 3.2., where it is 
discussed how the performances are dependent on the computational weight $w_i$ 
assigned to each particle.

\subsection{ Construction of the local essential tree}
A BH tree is constructed by each processor using the particles located in 
the processor subvolume. However, the local tree does not contain all 
of the information needed to perform the force calculation for the 
processor particles.
For these particles a subset of cells must be imported from the trees of the
 other processors according to the opening angle criterion applied to the 
remote cells. 
Each processor then receives a set of partial trees which are merged with the 
local tree  to construct a local essential tree \cite{du96}. The new local
tree contains 
all of the information with which the forces of the local particles can be 
consistently calculated.

The communications between processors of nodes from different trees imply
that in order to graft the imported cells onto the processor local trees 
it is 
necessary to adopt an efficient addressing scheme for the memory location of 
the nodes.
This is easily obtained if the construction of the local trees starts from a
root box of size $L$, common to all of the processors. Because of the ORB the
spatial domains of the particle processors will occupy a fraction $\sim  1/P$ 
of the computational volume $L^3$. This implies 
in the tree construction a small memory overhead. 
The main advantage is however that now the non-empty cells of
the local trees have the same position and size in all of the processors.
Each cell is then uniquely identified by a set of integers $\{j_1,j_2,...\}$,
with each integer ranging from $0$ to $7$ which identifies one of the $8$ 
subcells of the parent cell. These integers can be conveniently mapped 
onto a single integer word of maximum bit length $3k_{max}$, where 
$k_{max}$ is the maximum subdivision level of the tree. For a $64$ bit
key $k_{max} \le 21$.
This integer word represents the binary key of the cell.
When a cell of the tree is requested by a remote processor to construct its
local tree, the associated key is sent together with the mass, c.o.m. and 
multipole moments of the cell. The receiving processor then uses this key 
to quickly identify the cell location in the local tree and to add the new
cell to the local tree. A similar addressing scheme has been implemented, 
in their version of a parallel treecode, also 
by Miocchi \& Capuzzo-Dolcetta (2002).

The construction of the local essential trees is the most complicated 
part to be implemented in a parallel treecode. A simple translation of the 
logic of the BH opening criterion to a parallel treecode on a distributed
memory machine implies that each particle of each processor should apply 
the opening criterion to the tree nodes of the remaining $P-1$ processors.
This approach is clearly impractical because of the large communication cost 
needed to exchange particle positions or tree nodes between processors.
An efficient construction of the local essential tree is obtained instead as 
follows \cite{va02}.
This construction method is similar, with slight modifications, to the one 
described by Dubinski (1996).

i) Once the ORB has been completed and each processor has received the 
particle subset with spatial coordinates within its spatial domains,
the local trees are constructed according to BH in each of the 
processors $P_k$, where $k$ is a processor index ranging from $0$ to $P-1$.
 
ii a)  In a first version the  local essential trees were constructed as 
 in Dubinski (1996). 
A list of the cells of the processor's local tree is created, with the 
criterion that they must contain a number of particles $\simlt N_c (\sim 32)$.
This grouping scheme was introduced by Barnes (1990), in order to
significantly reduce the number of particle-cell applications of the opening
criterion necessary for constructing the interaction lists of the particles.
The opening angle criterion (1) or
(2) is applied between one of the grouped cells and the examined cells:
~$d$ is now the minimum distance between the cells c.o.m. and the grouped
cell. 
The obtained interaction list of cells by definition satisfies Eq. \ref{op}
or \ref{opn} for all the particles within the grouped cell.

\begin{table}
\label{tab:simpar}
\caption{Cosmological parameters of the simulations }   
\begin{center}
\begin{tabular}{ccccccc}
model& $\Omega_m$ & h &  $a_{fin}$ & $L(h^{-1}Mpc)$ &  $\sigma_8$ \\
\hline
 $\Lambda$CDM & 0.3 & 0.7 & 11  & 200 &  1.1 \\
 CDM & 1 & 0.5  &  40 &  11.11&  0.6 \\
\hline
\end{tabular}
\end{center}
\end{table}

This interaction list can then be used in the force 
calculation of all these particles. The speed-up depends on the distribution 
of the 
clustered particles and on machine hardware, Barnes (1990) suggests it can 
be a  factor of  $3$ to $5$.  However $N_c$ cannot be too large, otherwise
for a given grouped cell the generated interaction lists will contain
a number of cells largely in excess of those effectively needed.  
As a compromise, 
it is found $N_c \simlt 64$ \cite{ba90}.
For a multistep treecode the application of the grouping method for reducing
the cost of the list construction is not straightforward. 
This is because the number of active particles $N_{act}$
at a given time can be a small
fraction of the total particle number. 
In such cases, the advantages
 of having a unique interaction list for a small particle subset
 can be outweighed by the computational cost necessary for constructing it. 
The parallel treecode described here uses individual particle timesteps 
and this grouping method has not been implemented to construct the cell
interaction lists of the particle forces. 

After the list of grouped cells has been created, each processor
imports the root nodes necessary for the construction of the local essential 
tree from the other processors. 
These nodes are found by applying in the exporting processor the opening 
angle criterion to its local tree.
The list of cells and particles obtained is sent to the importing 
processor, where it is merged with the local tree at the corresponding nodes.
These nodes are identified using the binary keys of the imported nodes.
At the end of these steps each processor has the tree structure necessary
to compute the forces of the local particle distribution.

ii b) The communications between processors can be 
significantly reduced if one adopts the following criterion to construct the 
partial trees which will be exchanged between processors.
After the local trees have been constructed, each processor applies 
the opening angle criterion between the nodes of its local tree and the 
closest point of the volume of another processor $P_k$.
The partial tree obtained contains by definition all of the nodes of the local
processor necessary to evaluate the forces of the particles located in the
processor $P_k$. This procedure is performed at the same time by each
processor for all of the remaining $P-1$ processors. 
At the end, each processor has $P-1$ lists of nodes which are necessary for the 
construction of the local essential trees in the other processors.
The processor boundaries are determined during the ORB and are communicated 
between all of the processors after its completion.
Hence, the main advantage of this procedure is that all of the communications
between processors necessary for the construction of the local essential trees
are performed in a single all-to-all message-passing routine. 
The drawback of this scheme is the memory overhead, because each
processor imports from another processor a list of nodes in excess of those 
effectively needed to perform the force calculation, and also in excess of 
those which would have been imported with the previous procedure.
As a rule of thumb, it has been found that for $\theta=0.4$ a processor
with $N_p$ particles and $N_c$ cells imports $\sim N_p/8-N_p/4$ particles 
and $\sim N_c$ cells. The number of imported nodes is independent of 
the processor number. 
The value $\theta=0.4$ is a lower limit that guarantees
reasonable accuracy in the force evaluation in many simulations.
In the communication phase between processors, mass and position are imported 
for each particle, and the mass, c.o.m., quadrupole moment and the binary key 
are imported for each cell. 
The memory required by a single processor to construct the local essential 
tree is then approximately a factor $\sim2$ larger than that used in the 
implementation of the local tree. This memory requirement can be efficiently 
managed using dynamic allocation, and is not significantly larger than that 
required by other schemes used to construct the local essential tree
(e.g., Dubinski 1996).
 
\begin{figure}[h]
\vfill
\centerline{\mbox
{\epsfysize=10.0truecm\epsfxsize10.0truecm\epsffile{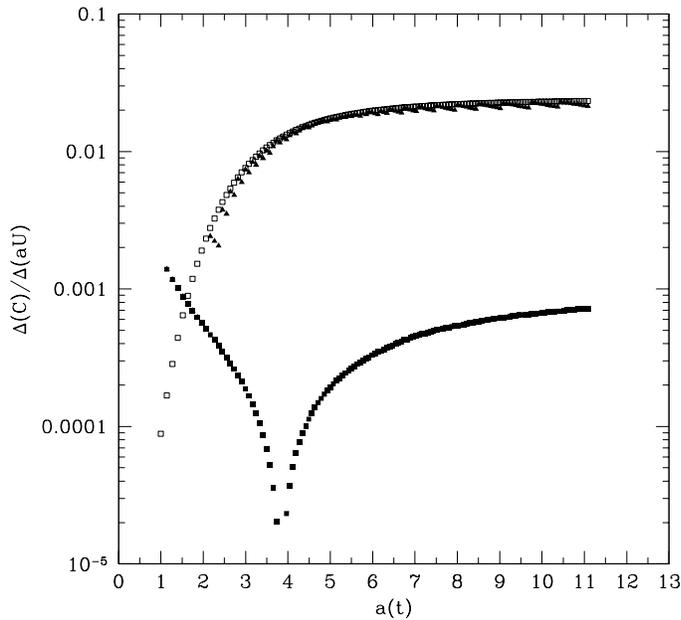}}}
\caption{
 The relative error $\Delta C/\Delta aU$ in the cosmic energy equation
is shown as a function of the expansion factor $a(t)$ for different
simulations. The chosen cosmological model is a flat CDM with a cosmological 
constant, $\omm=0.3$ and $h=0.7$. The simulations are performed in a
$200 h^{-1} Mpc$ comoving box with $84^3$ particles. Filled squares refer
to a simulation (S1L) with forces calculated using $\theta=0.4$ and quadrupole
 moments. Filled triangles correspond to a run (S2L) with the same simulation 
parameters but with $\theta=\theta(t)$ which is controlled by 
Eq. \ref{cvar} (open square).
Values of $\Delta C/\Delta aU$ for the corresponding parallel runs are not
displayed because they overlap with the plotted symbols.
              }
 \label{end}
\end{figure}

\subsection{Force calculation }
After the construction of the local essential trees has been 
completed, each processor proceeds asynchronously to calculate 
the forces of the active particles in its computational volume
up to the quadrupole order.  The code has incorporated periodic boundary 
conditions and comoving coordinates. Hence the forces obtained
from the interaction lists of the local essential trees must be 
corrected to take into account the contribution of the images.
\cite{da97,sp01}.
These correction terms are calculated before the simulation
 using the Ewald method \cite{he91}. 
The corrections are computed on a cubic mesh of
size $L$ with $50^3$ grid points and are stored in a file.
During the force computation, a linear interpolation is used to
calculate the correction terms  corresponding to the particle positions
from the stored values. 

   \begin{figure}[h]
  \vfill
\centerline{\mbox
{\epsfysize=10.0truecm\epsfxsize10.0truecm\epsffile{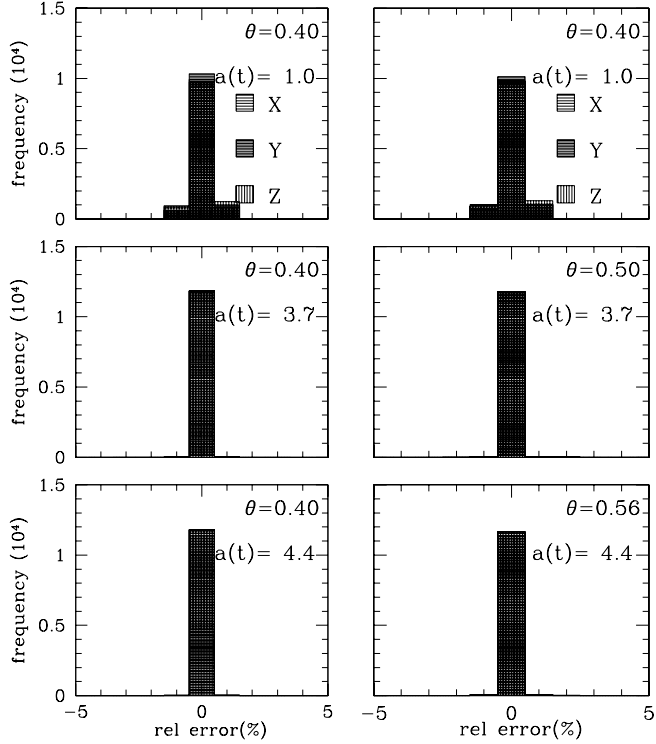}}}
\vspace{1cm}
\caption{
For two simulations with different values of $\theta$ 
the distribution of the relative errors in the force components is shown  
for several values of $a(t)$. 
The simulations are those of Fig. \ref{end}. Left column is 
for the run S1L
with $\theta=0.4$ and right column is for the run S2L with $\theta=\theta(t)$.
              }
         \label{herr}
   \end{figure}

In a cosmological simulation, the evaluation of the peculiar forces in 
the linear regime is subject to large relative errors. This is
because for a nearly homogeneous distribution, the net force acting on 
a particle is the result of the cancellation of the large partial forces
determined from the whole mass distribution.
From a set of test simulations Dav\'e, Dubinski \& Hernquist
(1997) found that in the 
linear regime, when 
$\theta=0.4$ and the cell moments are evaluated up to the quadrupole order, the 
relative errors in the forces are $\simlt 7\%$.
This problem is not present at late epochs, when the clustering of the 
simulation particles is highly evolved and even for $\theta \simeq1$
the relative errors in the forces are small ($\simlt 2\%$). 
This imposes in the simulation the necessity of varying $\theta$ according
to the clustering evolution, since the computational cost of evaluating
the forces with a small value of $\theta$ is wasted in the non-linear regime.
In this regime, the forces can be evaluated with an accuracy as good as that 
obtained in the linear regime, though using a higher value of $\theta$.

   \begin{figure}[h]
    \vfill
\centerline{\mbox
{\epsfysize=10.0truecm\epsfxsize10.0truecm\epsffile{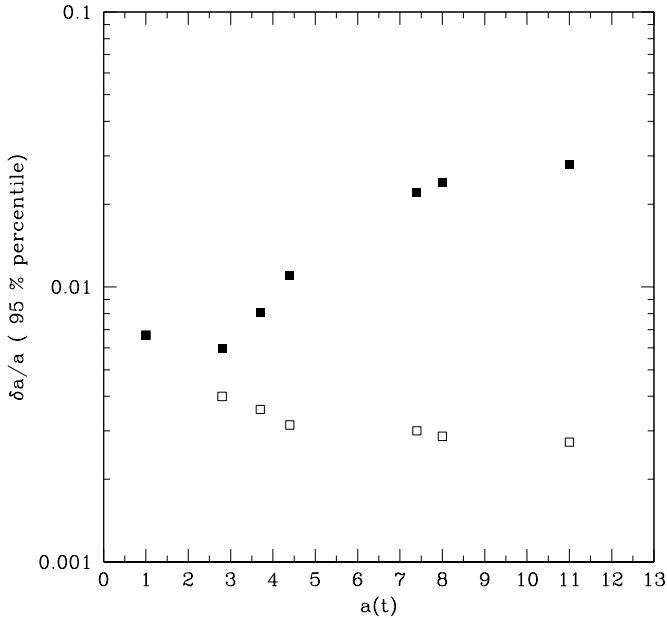}}}
\vspace{-1cm}
\caption{
Time evolution of the relative error $\delta a /a$ at the $95\%$ percentile of 
the cumulative distribution for simulations S1L
(open squares) and S2L (filled squares). 
The relative error is defined as  $\delta a /a \equiv |a-a_D|/|a_D|$, where 
$a_D$ is the particle acceleration evaluated in the direct summation limit.
            }
         \label{aerr}
   \end{figure}

After several tests it was found that a good criterion to control the 
value of $\theta(t)$ is that at any given simulation time $t$ the 
energy conservation must be satisfied with a specified  level of 
accuracy. The Lyzer-Irvine equation is 
\be
a^4 T +aU -\int U da =C,
\label{enc}
\ee
where $a=a(t)$ is the expansion factor, $T$ is the kinetic energy of the 
system, $U$ is the potential energy and $C$ is a constant. The accuracy
of the integration can be measured by the quantity
$err(t)=|\Delta (C)/\Delta(aU)|$, where $\Delta f$ denotes the change of 
$f$ with
respect to its initial value.
Fig. \ref{end} shows the time evolution of $err(t)$ versus 
the expansion factor for different test simulations. 
The cosmological model considered is a flat CDM model, with a vacuum energy 
density $\oml=0.7$, matter density parameter $\omm=0.3$ and Hubble constant 
$h=0.7$ in units of $100 Km sec^{-1} Mpc^{-1}$. 
Table 1 presents the main parameters of the model.
The power spectrum of the density fluctuations has been 
normalized so that the r.m.s. mass fluctuation in a sphere of 
radius $8 h^{-1} Mpc$ takes the value $\sigma_8=1$ 
at the present epoch, $a(t)=a_{fin}=11$. The simulations are run in an 
 $L=200 h^{-1}Mpc$ comoving box  with $N_p=84^3$ particles.
The gravitational softening parameter of the particles is set to
$\varepsilon_g=L/10N_p^{1/3}$.
The filled squares of Fig. \ref{end} show $err(t)$ 
 for a simulation with $\theta=const=0.4$. This simulation (S1L) was
performed using the serial version of the code.
With this value of $\theta$ one has $err(t)\simlt 10^{-3}$ even in
 non-linear regimes, when $a(t)$ approaches its final value. 
For several values of $a(t)$ the distribution of the relative root mean 
square errors in the 
force components of a particle subset ($\sim 5\%N_p$) of the S1L run 
are displayed
 in 
Fig. \ref{herr}
(left column). The reference forces are calculated from the same 
particle distribution using $\theta=0.01$.
It has been found that a similar distribution in the relative errors is 
obtained if during the 
simulation $\theta=\theta(t)$ increases with time, provided that its value
never exceeds an upper limit
 implicitly defined for $\sigma_8 \geq 0.2$ 
by the constraint
\be
\Delta C /\Delta (aU) \leq 0.025 /[1+(0.4/\sigma_8)^{3}]^{1.7}.
\label{cvar}
\ee

An additional constraint sets an upper limit $\theta \leq 0.9$. 
A simulation was performed (S2L) with the same input parameters of the 
run S1L, but with the value of $\theta$ now implicitly controlled by Eq. 
\ref{cvar}. 
The level of energy conservation given by the constraint (5) 
is shown in correspondence of run S2L in Fig. \ref{end} (open squares). 
The relative errors in the force components of run S2L are plotted
 in the  right column of Fig. \ref{herr} for the same
particle subset of the run S1L.
A more quantitative comparison between the two distributions can be obtained
from Fig. \ref{aerr}. The time evolution of the 
 $95\%$ percentile of the cumulative 
distribution of the relative acceleration errors is plotted for the two
runs  S1L (open squares) and S2L (filled squares). 
The relative acceleration error  is defined as 
$\delta a /a \equiv |a-a_D|/|a_D|$, here $a_D$ is the particle acceleration 
evaluated in the direct summation limit.
The results show that for simulation S2L the relative errors in the 
forces are always below $\sim 3\%$, an upper limit which has been found 
to yield reasonable accuracy in various tests \cite{da97,sp01,mi02}.

Therefore, the criterion (5) can be profitably used to constrain the
value of $\theta(t)$ according to the clustering evolution, and at 
the same time to maintain the relative errors in the forces below
a fixed threshold ($ \simlt 3\%$). This allows a substantial increase 
in the code performances. The computational cost of evaluating the forces 
depends on $\theta$ and for the considered runs 
 at $a(t)=11$  it
is significantly reduced by a factor of $10$ to $20$ when $\theta$ 
is increased from $0.4$ to $\sim0.9$.

\subsection{Multiple timesteps and particle update}
After the force calculation is complete, particle velocities and positions are 
updated in each processor. In the individual timestep scheme \cite{hk89} the 
particle timestep $\Delta t_i$ of particle $i$ is determined according to 
several criteria.
The first is important at early  epochs and requires that  
\be 
\Delta t_i \leq \Delta t_{exp} =0.03 ~2/3~H(t),
\label{dt1}
\ee
where $H(t)$ is the Hubble parameter at the simulation time $t$.
The other two criteria are 
\begin{equation}
\Delta t_i \leq 0.3 ( \varepsilon_i~ a^3(t) /g_i)^{1/2}
\label{dt2}
\end{equation}
\begin{equation}
\Delta t_i \leq 0.3 ( \varepsilon_i/v_i)~,
\label{dt3}
\end{equation}
where $\varepsilon_i$ is the comoving gravitational softening parameter of 
the particle
 $i$, $g_i$ is the peculiar acceleration and $v_i$ its peculiar velocity.
These criteria are similar to those adopted by Dav\'e, Dubinski \& Hernquist
(1997).
Particle timsteps are constrained to take the values 
$\Delta t_j = \Delta t_o/ 2^{j}$, where $j \geq0$ is an integer. 
For the particle $i$ the timestep $\Delta t_i$, which satisfies the above
criteria, takes the value $\Delta t_{j_i}$ such that 
$\Delta t_{j_i} \leq \Delta t_i$.

 At the beginning of the integration $t=t_{in}$, the forces are evaluated for 
all of the 
particles and their positions are advanced by half of the initial timestep,
which is common to all the particles.  
In this integration scheme, the forces are evaluated at later times $t>t_{in}$ 
only for those particles for which it is necessary to maintain the second-order
accuracy of the leapfrog integrator.
Particles can change their time bins $\Delta t_j$ and their positions must 
be corrected to preserve time centering. The transformations necessary 
to perform these corrections in comoving coordinates are given in 
appendix A.
The particle positions are advanced using the smallest timestep 
$\Delta t_{min}$, as determined by the above constraints.
In the parallel implementation, each processor determines the individual
particle timesteps and the smallest timestep $\Delta t_{min}^{(pr)}$
of its particle subset, $\Delta t_{min}$ is then the smallest of these 
 $\Delta t_{min}^{(pr)}$
and is used by all the processors.

\begin{table}
\label{tab:summ}
\caption{Summary of the simulations }
\begin{center}
\begin{tabular}{ccccccc}
 model &  $N_p$~$^a$ & $\theta$~$^b$ & procs$^c$ & run$^d$ \\ 
\hline
$\Lambda$CDM &  $84^3$ & 0.4 &  & $S1L$ & \\
 $\Lambda$CDM &   $84^3$ &   var&  & $S2L$ & \\
 CDM &  $32^3$ & 0.4 & P  & $S3C$ &\\
 CDM &  $\sim 35,127$ & 0.4 & P  & $S4C$ &\\
 CDM &  $140,608 $ & 0.4 &   & $S5C$&$(S6C:SPH)$ \\
\hline
\end{tabular}
\end{center}
\vglue 0.2truecm
\par\noindent
{\it Note } 
{\small 
$^a$ : number of particles, for parallel runs is the average number of 
particles per processor. 
$^b$ : the symbol $var$ denotes $\theta=\theta(t)$, 
where $\theta$ is implicitly defined  according to Eq. \ref{cvar}. 
$^c$ procs is the number of processors in the parallel runs.  
$^d$ : serial runs are indicated with an $S$, the parallel
runs have the same simulation label of the corresponding serial runs but
with a subscript added to indicate the processor number $P$. 
The notation $x$ means that the total particle number of the simulation 
is that of the serial run times the number of processors. 
The simulation S6C is hydrodynamic (SPH), the cosmological and simulation 
parameters are the same of the run S5C, but with a baryonic fraction
$\omb=0.05$. The simulation includes radiative cooling and is performed
with $N_p=140,608$ dark matter and gas particles.
}
\end{table}

After the particle positions have been updated, it may happen that a
fraction of the particles assigned to a given processor has escaped
the processor subvolume.
At each timestep, the particles which are not located within the original
processor boundaries are moved between the processors.
 In principle, the computational cost of locating the processor to which 
the escaped particle belongs
scales as the processor number $P$ ( Dav\'e, Dubinski \& Hernquist
1997, sect. 5.1). 
However, if during the ORB the processors are partitioned according to the 
procedure described in sect. 2.1, the final processor ordering 
makes it possible to reduce to $\sim \log_2 P$ the number of positional tests 
 of the particle. This is not a significant improvement
in pure gravity simulations, where the fraction of particles  which
leave a processor at each step is small ($ \sim 5\% $), but it is important 
in a future implementation of the code which will incorporate 
smoothed particle hydrodynamics (SPH; Dav\'e, Dubinski \& Hernquist 1997, 
Springel, Yoshida \&  White 2001, Springel \& Hernquist 2002).
In such a scheme, gas properties of a 
particle are estimated by averaging over a number of neighbors of the particle; 
 in a parallel implementation, an efficient location of the particle 
neighbors located in the other processors is important in order to improve the
code performances.

\section{Performances}
The treecode described here uses the BH tree algorithm to compute the 
gravitational forces. The implementation of this algorithm is identical 
to that of Dubinski (1996) and Dav\'e, Dubinski \& Hernquist (1997). 
The dependence of the errors in the force evaluations on a number of
input parameters has been discussed previously by these authors,
therefore an error analysis of the forces will not be presented here.
 It is worth noticing that, in the parallel treecode described here, the
tree construction in each processor is such that the gravitational force 
acting on a particle  is identical to that computed by the serial version 
of the code. For a given particle distribution, a comparison of the values of 
the forces obtained by the parallel version with the corresponding ones 
of the serial version can therefore be used to control that there are not 
bugs in the parallel routines which perform the construction of the local
 essential trees.

\begin{figure}[h]
\vfill
\vspace{2.5cm}
\centerline{\mbox
{\epsfysize=10.0truecm\epsfxsize=10.0truecm\epsffile{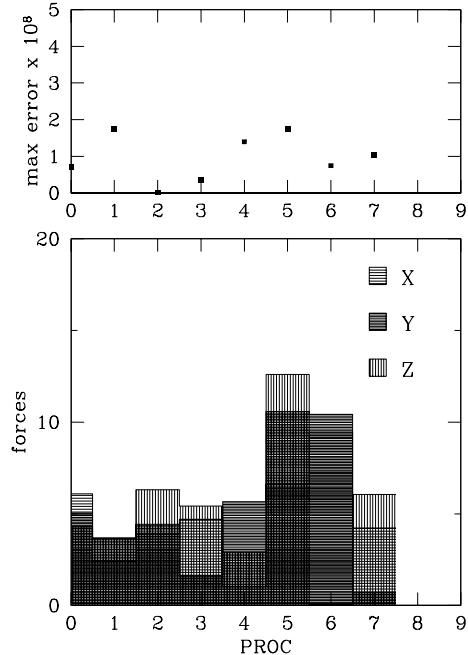}}}
\vspace{-3cm}
\caption{ For the simulation $S3C^x\_8$ the upper panel shows the maximum 
absolute errors in the force components evaluated by each of the $P=8$ 
processors at $a(t)=1$. Reference forces have 
been calculated with a treecode
algorithm applied to the whole distribution of $32^3P$ particles.
The opening angle parameter is $\theta=0.4$ and cell moments are 
calculated up to quadrupole order.
The bottom panel shows the corresponding values of the forces.
}
  \label{acmx}
\end{figure}

The results of this test are shown in Fig. \ref{acmx} for a configuration with
$P=8$ processors. The initial positions of $32^3 P $ particles in a 
$L=11.11 h^{-1}Mpc$ comoving box have been perturbed according to a CDM model 
with $\omm=1$, $h=0.5$ and power spectrum normalization $\sigma_8=0.7$ at the
present epoch (simulation $S3C^x\_8$ of Table 2). 
This epoch corresponds to a value $a_{fin}=40$ of the expansion factor and
 particle forces have been evaluated at $a(t)=1$.
This is because in linear regimes the computation of the 
forces for a tree code is subject to large relative errors (see sect. 2.3).
For a comparison of the forces obtained with different tree codes 
from the
same particle distribution, the choice $a(t)=1$ then corresponds to the most 
severe particle configuration to be used with these parameters.
Forces have been calculated with an angle parameter $\theta=0.4$ and 
quadrupole corrections. To compute parallel forces the particle distribution 
was split among the $P$ processors.
The results plotted in Fig. \ref{acmx} show that the maximum relative difference
between the forces computed by the treecode and its
parallel version is $\sim 10^{-8}$. Analogous results were obtained 
for a configuration with $P=64$ processors.
 These small differences are presumably due to round-off errors, 
which lead in the two code versions to differences of the same order in the
values of the moments of certain cells. 
These differences follow because these cells have spatial volumes which 
occupy two or more processor domains.
In the parallel treecode the moments of these cells are then evaluated by 
summing partial contributions in a different order from that of the serial
version.
For the simulations of Fig. \ref{end} the values of $\Delta C/\Delta aU$ of the 
corresponding 
parallel runs are not shown because they practically overlap with those 
of the serial runs.

\subsection{Scalability}
The computational speed of the code is defined as the particle number divided
by the elapsed CPU wall-clock time \tc spent in the force computation of the 
particles. This definition includes also the treewalk necessary for
constructing the interaction lists.
 For a specified accuracy and particle distribution, 
the CPU time \tc 
of a parallel treecode with maximum theoretical efficiency 
is a fraction $1/P$ of that of the serial code.

\begin{figure}[h]
\vfill
\vspace{4cm}
\centerline{\mbox{
\epsfysize=10.0truecm\epsfxsize=14.0truecm\epsffile{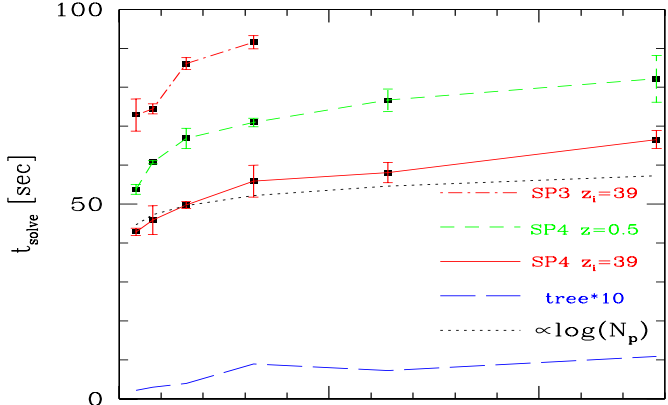}}}
    \vspace{-5cm}
\caption{
The average elapsed CPU time \tc spent in the force evaluation 
 as a function of the number of processors $P$. The value of the 
opening angle parameter is $\theta=0.4$, quadrupole moments are taken into
account.
The error bars are the dispersions over the  $P$ processors.
The initial particle
configuration is found from a uniform distribution perturbed according to
a CDM power spectrum at $z_i=39$. 
The tests were performed on the IBM SP3 ($P\leq32$) and SP4 machines 
($P\leq128$).
The dash-dot line  shows the values of \tc for the tests $S3C^x\_P$ on the 
SP3, continuous line refers to the tests $S4C\_P$ on the SP4 (see Table 2).   
 The total particle number of the tests $S4C\_P$ is given in Table 3.
The short dash line connects the values of \tc 
 obtained from a set of clustered distributions  
statistically equivalent to that of the runs $S4C\_P$ at $z=0.5$   
(see text).
The long dashed line is the CPU time
for constructing the tree scaled up by a factor 10. 
The dotted line is the expected \tc
from the ideal scaling relation $\propto log(N_p)$.
}
         \label{tso}
\end{figure}

The scalability of the parallel treecode was tested by measuring \tc 
using a different number of processors $P$.
For a configuration of $P$ processors \tc is defined as the average
of the values of the individual processors.
The tests were performed on the IBM SP3 and SP4 machines hosted  by
CINECA (Bologna, Italy). In its present configuration the IBM SP4 has
$16 Gb$ of RAM memory available on each 8 processor node. 
The cosmological model is the CDM model previously described. 
 The simulation parameters of the runs, $S3C\_P$ and $S4C\_P$, 
are given in Tables 2 \& 3. 
For these runs  the number of particles scales
linearly with $P$. This dependence of $N_p$ on $P$ has been chosen in
order to compare the force solving CPU time \tc consistently with
the one necessary for the construction of the local essential tree.
If $N_p$ is kept fixed and $P$ becomes very large the communication time 
necessary for the construction of the local essential trees is expected to be 
dominant over the force computation and thus code performances will be
degraded. On the other hand, this configuration is of scarce practical 
interest because parallel runs are expected to be performed 
with a number of particles per processor as high as possible.
With the choice $N_p/P=const$, the CPU time \tc scales ideally 
as $t_{ideal}\propto \log N_p \propto \log P$. The results are shown in 
Fig. \ref{tso},
where $t_{solve}(P)$ is plotted up to $P=128$ (continuous line) for the tests on
the SP4 and up to $P=32$ (dash-dot line) for tests on the SP3.
The dotted line shows the ideal scaling relation. In the large $P$ limit 
\tc is approximately $\sim10\%$ higher than the ideal scaling relation.
This is most probably due to cache effects of the machine which arise 
 when $N_p \simgt 10^5$ during the tree descent necessary to calculate the 
forces.
The computational speed  at $P=16$ is 
$ \sim 32^3P/t_{solve} \sim 400P ~part/sec$ on the IBM SP3, 
and $\sim 700 P~part/sec$ on the SP4. 
This is valid for $\theta=0.4$. If $\theta$ is increased, the particle 
interaction list will have a smaller number of terms and \tc will be
smaller. It has been found that $t_{solve}(\theta)$ is well  approximated by
 $t_{solve}(\theta \leq1)\propto10^{-5\theta/3}$.
If $\theta=1$ the CPU times of Fig. \ref{tso} will then be reduced by a factor 
$\sim 10$. 

\begin{table}
\label{tab:paral}
\caption{ Number of particles of the parallel runs $S4C\_P$}
\begin{center}
\begin{tabular}{ccc}
 P~$^a$ &  $N_p$ & $N_p/P\sim$~$^b$ \\ 
\hline
 4 & 140608  &     35152  \\
8  & 287496   &     35937 \\
 16 & 551368   &     34460 \\
32 & 1124864    &  35152 \\
64 & 2197000    &   34328 \\
132 & 4574296  &  35737  \\
\hline
\end{tabular}
\end{center}
\vglue 0.2truecm
\par\noindent
{\it Note } 
{\small 
$^a$ : number of processors. 
$^b$ : average number of particles per processor.}
\end{table}

An important result is the time $t_{tree}$ required to construct the local 
essential trees.
The dashed line of Fig. \ref{tso} shows that this time is always a small 
fraction ($\simlt 2 \%$) of the time required to compute the gravity.
It is worth noticing that the communication part is efficiently handled by
the all-to-all routine and the corresponding time is a negligible fraction of
$t_{tree}$. 
This is valid if the accelerations are calculated for the whole particle set.
In the parallel implementation of the multistep treecode described here, the 
local essential trees are constructed at each timestep and the corresponding 
computational cost is approximately constant, whereas the force solving 
CPU time \tc depends on the number of active particles at the given timestep.
In such cases it may happen that \tr $\geq$ \tc. At the end of a large timestep
$\Delta t_0$ the total tree construction time \tr$(\Delta t_0)$ is defined as 
the sum of the partial times \tr$(\Delta t_j)$ which have been recorded 
between $t_n= n \Delta t_0$ and $t_{n-1}$; a similar definition holds also 
for \tc$(\Delta t_0)$. The ratio \tr$(\Delta t_0)/$\tc$(\Delta t_0)$  
depends on the particle timesteps distribution and the lower limit 
$\Delta t_{min}$. For the tests discussed here 
$\Delta t_{min} = \Delta t_0/32$ is the minimum considered 
value of $\Delta t_{min}$. 
According to this constraint it was found that for the considered 
parallel runs with an evolved clustering state, the ratio 
 \tr$(\Delta t_0)/$\tc$(\Delta t_0)$  is always small ($\simlt 10 \%$).

In order to assess how \tp scales 
with $P$ in the strong clustering regime
 it would be necessary to perform a set of 
simulations down to
$z=0$ using a different number of processors for the same model. In the 
large $P$ limit
this requires a considerable amount of computational resources.
Because the aim of the tests is to obtain an estimate of \tp 
for a clustered distribution, then \tp can be measured
according to the following procedure.
A serial run (S5C) was performed using $N_p=140,608$ particles, with the
same cosmological parameters and initial conditions of the tests $S4C\_P$.
The run was evolved from $z_i=39$ to $z=0$.
The  maximum timestep takes the value
$\Delta t_0=t_{fin}/424$, and $\Delta t_{min}=\Delta t_0/32$.  
Particles timesteps are constrained according to Eqs. \ref{dt1},\ref{dt2}
and \ref{dt3}.
The redshift $z=0.5$ corresponds to a simulation time 
 $t_n=n \Delta t_0=228 \Delta t_0$.
At this redshift the particle distribution was divided within the 
computational volume $L^3$ using the ORB procedure and particle computational
weights calculated at the previous step. The four subvolumes and their 
corresponding particle distributions were given as initial conditions to 
the parallel treecode with $P=4$ processors.
The average number of particle per processor is $\sim 35,000$ and is that of 
$S4C\_P$. The first point of the CPU time \tp at $z=0.5$ is then measured 
as in the previous test at $z_i=39$.
For a larger number of processors (up to $P=32$), clustered distributions 
statistically equivalent to those of the tests $S4C\_P$ can be obtained 
as follows. The four subvolumes of the test with $P=4$ processors and 
$N_p(P)$ particles have 
lower bottom corners with coordinates 
$xb_p^m$, sizes $l_p^m$ and volumes $V_p =l_p^1l_p^2l_p^3$, where $p$ is 
the processor index and $m$ is the axis index.
The subvolumes are divided along a cartesian axis $m_1$
into two new separate equal volumes $V_p/2$, the particles which belong to the 
original subvolumes
have now half the original mass and coordinates $x_i^{m_1} \rightarrow 
xb_p^{m_1} +(x_i^{m_1}-xb_p^{m_1})/2=x_i^{m_1}(1)$. In each of the original
subvolumes the particle distribution 
is then replicated creating a new set of $N_p(P)$ particles with coordinates 
$x_i^{m_1}(2)$ given by
$x_i^{m_1}(2)=x_i^{m_1}(1)+l_p^{m_1}/2$ and all of the other variables 
left unchanged. This procedure can then be repeated along the other two axes
to generate up to 8 subdivisions from each of the four subvolumes
used for the test with $P=4$. To measure \tp at $z=0.5$  for $P=64(P=128)$
this splitting procedure is first applied along one (two) cartesian 
axis to the whole particle distribution within the computational 
volume $L^3$. The new distributions are divided among 8 (16) subvolumes and the 
procedure described above is used to generate for each of the subvolumes 
8 new subdivisions.
After the whole procedure has been completed, the final particle 
distributions within the original volume $L^3$ are then used as initial 
conditions for measuring \tp using the parallel treecode with $P=64$ or 
$P=128$.  The obtained values of \tp range from 
$P=4$ up to $P=128$ and are displayed in Fig. \ref{tso} (short-dash line).
These values are supposed to give a fair estimate of what would be 
those obtained running the tests $S4C\_P$ down to $z=0.5$.
The results show that $t_{solve}(P,z=0.5)$ scales with $P$ as 
 $t_{solve}(P,z_i=39)$ in the linear regime; this demonstrates that the 
overall scalability of 
the code does not depend upon the degree of clustering of the particle 
distribution.

A comparison of the code performances with those of other authors 
is difficult because of different algorithms, particle distributions and
machines. The plots of Fig. \ref{tso} show that the values of \tc on the
IBM SP4 are only a factor $\sim 2$ smaller than the corresponding ones  
measured on the IBM SP3, though the SP4 has a clock-rate which is 
$\sim 3.5$ ($1.3 GHz$) faster than that of the SP3 ($375 MHz$).
These relatively poor code performances on the SP4 must be attributed 
to the different hardware architectures and size of the
memory caches in the two machines, see Genty et al. (2002) for the relative 
documentation on the SP4.
In order to compare code performances with those illustrated in other papers
it is more appropriate to refer to the values of \tc measured with the IBM SP3.
An educated guess that the code has performances fairly comparable
 with those of the Springel, Yoshida \&  White (2001) code
 is given by Fig. 12 of their paper.
 In this figure the gravity speed as a function of the processor number is
shown for a cosmological hydrodynamic SPH simulation in a comoving box size
of $50 h^{-1} Mpc$ with $32^3$ dark matter particles and $32^3$ gas particles.
The cosmology is given by a \La~ model with $\omm=0.3$ and $h=0.7$. 
The simulations were evolved from an initial redshift $z_i=10$.
The plotted speeds were measured on a CRAY T3E ($300MHz$ clock).
From Fig. 12 the computational speed of gravity for $P=32$ is 
$\sim 7500 part/sec$.
The cosmological model is not that adopted in the tests of Fig. \ref{tso}, but  
at early redshifts the computational cost of the gravity force calculation
is  not strongly dependent on the assumed model.
For $P=32$ and $\theta=0.4$ the results of Fig. \ref{tso} give a gravity speed 
of $\sim 12 \cdot 10^3 part/sec$ on the IBM SP3 for a CDM model at $z_i=39$.
The measured speed must be reduced by $\sim 20\%$ to take into account the 
higher clock-rate of the IBM SP3 ($375 MHz$). 
The final value 
($\sim 9500 part/sec$) is similar to 
the one obtained by Springel, Yoshida \&  White (2001). 
It must be stressed that the timings shown in Fig. \ref{tso} are the worst 
case to be considered for a comparison with the gravity speeds reported in Fig. 12 of Springel et al. (2001).
This is because the clustering evolution is expected to allow a decreasing
value of $\theta(t)$ and thus a corresponding increase in the computational 
speed. The CPU times \tc of Fig. \ref{tso} refer to the gravitational 
 force calculation
 with $\theta=0.4$ of a particle distribution at high redshifts, whereas
the speeds in Fig. 12 of Springel, Yoshida \&  White (2001) 
are defined over the entire range of 
the simulation time.
The particle forces of the tests shown in Fig. \ref{tso} were evaluated 
applying the improved BH criterion (2) with $\theta=0.4$ and quadrupole 
moments to the particle distributions. 
For $P=4$ the corresponding particle interaction lists have at 
$z_i=39$ a mean number of terms 
$<w>\sim 1,000$. The distribution of the relative 
acceleration errors  shows that  
$95\%$ of the particles have $\delta a/a \simlt 10^{-2}$.
These results can be compared with those displayed in Fig. 7 of 
Springel, Yoshida \&  White (2001), where the $95\%$ percentile 
of the cumulative distribution of the relative force errors 
 $\delta a/a (< 95\%)$
is plotted versus $<w>$ in a $32^3$ cosmological simulation at $z_i=25$.
The plots are for different opening criteria.  
At $\theta=0.4$ the new opening criterion of 
Springel, Yoshida \&  White (2001) gives a comparable accuracy with 
approximately the 
same number of terms. 

Another comparison can be made with the parallel treecode developed 
by Miocchi \& Capuzzo-Dolcetta (2002). 
The authors computed gravity speeds on a CRAY T3E for a 
Plummer distribution
with $N_p=128,000$ particles.
Fig. 7 of their paper shows $\delta a/a (< 90\%)$ versus 
the average work $<w>$ per particle.
The forces were calculated using the BH criterion with quadrupole moments.
The maximum value of $<w>$ in the figure is $\sim 1,000$, and it corresponds
to $\delta a/a (< 90\%) \sim 10^{-3}$. 
For a configuration of $P=32$ processors 
this value of $\delta a/a (< 90\%)$  
gives a gravity speed of about $\sim 10^4 part/sec$ (Fig. 5 of their paper).
Taking into account the different clock rates of the machines used in the 
tests, this value of the speed is approximately that obtained here on the IBM
SP3 with the same number of processors and an interaction list of nearly equal
length.
Finally, is is worth noticing that the ratio \tr/\tc is significantly 
lower than that obtained in other parallel versions of a treecode
\cite{du96,sp01}.

\begin{figure}[h]
\vspace{-3cm}
\centerline{\mbox
{\epsfysize=12.0truecm\epsfxsize=10truecm\epsffile{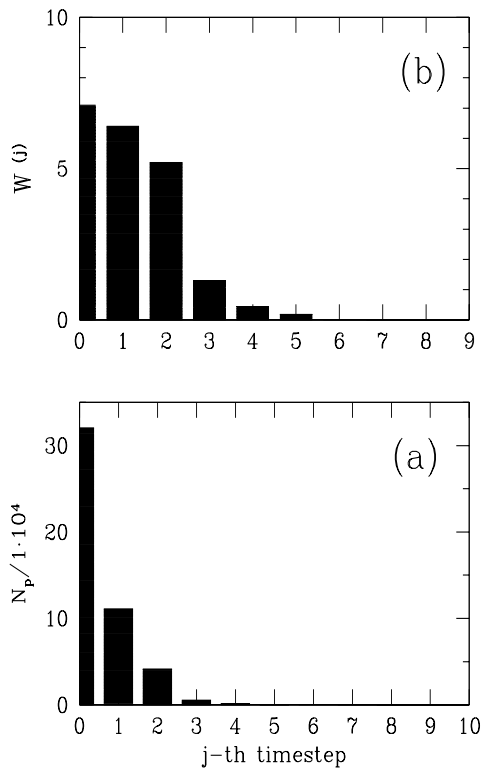}}}
\caption{ {\bf{(a):}} Number of particles with timesteps 
$\Delta t_j = \Delta t_0/2^j$
at the end of a macrostep $\Delta t_0$ for the serial run S6C.
The number of particles is $479,445$,  
$\Delta t_0=t_{fin}/424$, $\Delta t_{min}=\Delta t_0/32$ and 
simulation time $t_n=n \Delta t_0=228 \Delta t_0$.
The simulation is hydrodynamic according to the SPH method,
and includes radiative cooling and star formation.
The number of particles includes also the star particles which
have been formed since the initial simulation time.
{\bf {(b):}} For the same particle distribution  
the corresponding computational loads 
$W^{(j)}=\sum _{i \in \Delta t_j} W_i$ are shown. 
The summation is over all the 
particles $i$ with timesteps $\Delta t_j$. 
       }
         \label{whist}
\end{figure}

\subsection{Load balancing}
An important characteristic of a treecode is load balancing.
An ideal code should have the computational work divided evenly 
between the processors. This is not always possible and code performances
will be degraded when the load imbalance  between the processors becomes
large. 
At any point of the code where  
synchronous  communications are necessary 
there will be $P-1$ processors waiting for the 
most heavily loaded one to complete its computational work.
Load balancing can then be measured as
\be
L= \frac{1}{P} \sum_p 1 -(t_{max}-t_p)/t_{max},
\label{lbal}
\ee
where $t_p$ is the CPU time spent by the processor $p$ to complete a
procedure and $t_{max}$ is the maximum of the times $t_p$.
A treecode spends most of the CPU time in computing gravitational forces, and  
so it is essential to have good load balancing ($\simgt 90\%)$ with the 
gravity routine. 
As already outlined in sect. 2.1 this task is not obviously achieved
with a multistep treecode.
The number of active particles $N_{act}$ between the simulation times $t_n$ 
and $t_n+ \Delta t_0$ can vary wildly as a function of the current 
simulation time $t_n^{(k)}$. This is defined 
$k$ steps after $t_n$ as $t_n^{(k)}=t_n +\sum_{j=1}^{j=k} 
\Delta t_j$, the summation is at $t > t_n$ over the past timesteps $\Delta t_j$.
 
At a certain step the ORB procedure described
in sect. 2.1 can be used to obtain load balancing, but at later steps the
unbalancing can be substantial. 
This problem has prompted some authors to consider more complicated 
approaches \cite{sp01,mi02}.
Here a simpler route is followed starting from the observation
that in a multistep integration scheme, a better measure of the 
computational work done by each particle $i$ is given by 
$W_i =\sum _k w_i^{(k)}$, where $w^{(k)}_i \propto N_{OP}(i)$ is the 
number of floating point operations of particle $i$ necessary to calculate the 
gravitational forces of the particle at the simulation time $t_n^{(k)}$. 
If the particle $i$ is not active at $t_n^{(k)}$ , $w^{(k)}_i=0$.
The summation is over the steps 
between $t_n$ and $t_n + \Delta t_0$, the weights $W_i$ are now used  at
each large step $\Delta t_0$ to subdivide the computational volume
according to the ORB procedure.
Theoretically, this weighting scheme does not guarantee a perfect 
load balance, nonetheless it has been found to yield satisfactory
results ($L \simgt 90 \%$) in many typical applications.
The reason lies in the shape of the distribution function $F(\Delta t_i)$ 
of the particle timesteps $\Delta t_i$, for a simulation with an evolved 
clustering state. 
The number of particles with 
timesteps in the interval $\Delta t_j, \Delta t_j+ \Delta t$ is
given by $n_j=F \Delta t$.   
The particle timesteps are determined according to the criteria 
defined in sect. 2.4. Another parameter which determines the shape of
the distribution function is the minimum timestep $\Delta t_{min}$.
The optimal choice for $\Delta t_{min}$ requires that 
 the number of particles  
of the binned distribution $n_j$ in the last time bin should be  
 a small fraction of the total particle number, as a rule of 
thumb it is found  $N_{opt}\sim 5 \%N_p$.
 The particles in the last time bin are $n_{jmax}$ with timesteps 
$\Delta t_{min}=\Delta t_{jmax}=\Delta t_0/2^{jmax}$.
If $n_{jmax}$ is much higher than the threshold $N_{opt}$ it means that
the integration scheme is not accurate, 
for a specified set of constraints on the particle timesteps, 
because there are many 
simulation particles with $\Delta t=\Delta t_{min}$. On the 
other hand, if $n_0 \sim N_p$, it would imply that the computational work of 
the code is wasted for the required accuracy.

The binned distribution $n_j$ was measured for a clustered 
particle distribution originated from a serial (S6C) test 
simulation. This simulation is similar to the serial run S5C of sect. 3.1,
but incorporates now hydrodynamics according to the SPH scheme.
The cosmological parameters are the same as S5C, but with a baryonic
density parameter $\omb=0.05$. These values are the same simulation 
parameters used by  Dav\'e, Dubinski \& Hernquist (1997, sect. 4.3) for 
testing their 
parallel SPH treecode. The run was performed from $z_i=39$ down to
$z=0$ using an equal number of $140,768$ gas and dark matter particles.
The simulation includes radiative cooling, an ionizing UV background and
gas particles in cold high density regions are subject to star formation.
For more details see, e.g., Valdarnini (2002b).
The maximum timestep takes the value
$\Delta t_o=t_{fin}/424$, and  $\Delta t_{min}=\Delta t_0/32$  is the 
minimum allowed timestep of dark matter particles according to the criteria 
defined in sect. 2.3; gas particles have their timesteps additionally 
constrained by the Courant condition \cite{hk89}. 
At the simulation time $t_n=n~ \Delta t_0=228~ \Delta t_0$, the particle 
distribution of the simulation was used, following the splitting procedure
described in sect. 3.1, for generating the initial conditions for the parallel
simulations. The parallel runs are purely collisionless and 
 gas particles were treated as  dark matter particles. 
These simulations were performed to investigate the load 
balancing efficiency of the parallel treecode.
According to the previous discussion, the load balancing parameter
$L$ depends on the shape of the distribution function of the particle
timesteps. For this reason it was decided to generate initial 
conditions for the parallel runs from an SPH simulation. Hence,
the measured performances are indicative of those that would be obtained 
if SPH were to be implemented in the parallel code.

   \begin{figure}[h]
\vfill
\vspace{3cm}
\centerline
{\mbox{\epsfysize=10.0truecm\epsfxsize=14cm\epsffile{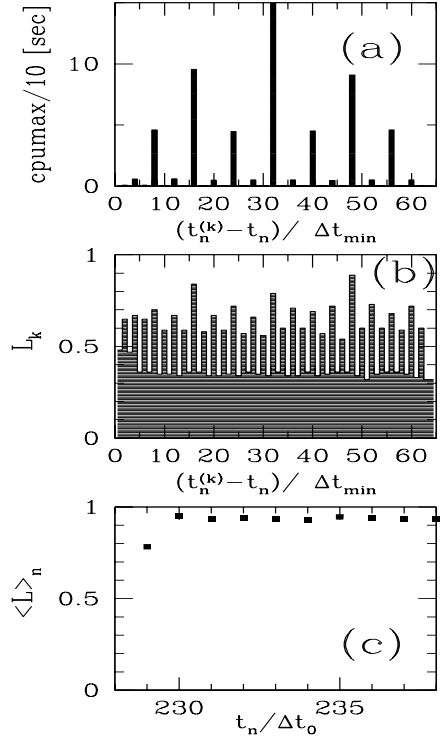}}}
\caption{ 
The load balancing scheme is tested for a parallel run with $P=4$ processors.
The initial conditions have been obtained from the clustered particle 
distribution of simulation S6C (see Fig. \ref{whist}).  
{\bf{(a):}} 
The top panel shows, between $t_n$ and $t_{n+1}$, the maximum of the CPU times 
of the $P$  processors at the simulation time $t_n^{(k)}$, $n=228$.
The corresponding load balancing $L_k$ is plotted in the mid
panel {\bf {(b)}}.
The bottom panel shows  the load balancing $<L>_n$ at the end of each macrostep
$\Delta t_0$. The validity of the weighting scheme is shown by the first
point, where the ORB procedure has been performed setting $w_{i}=const$.
}
 \label{loadh}
\end{figure}

The distribution  $n_j$ of the simulation S6C is shown 
 as a function of the particle timesteps $\Delta t_j$ in Fig. \ref{whist}a,
at the simulation time $t_n=n~ \Delta t_0=228~ \Delta t_0$.
The corresponding distribution of particle computational loads $W_i$ is
shown in panel (b). The plotted distribution is $W^{(j)}=\sum _{i \in
\Delta t_j} W_i$, the summation being over all of the particles $i$ with 
timesteps $\Delta t_j$. 
About $\sim 90\%$ of the particles are in the first three time bins. It can
be seen that for these bins the variations in the load distribution are 
within $\sim 20\%$. For example, the number of particles $n_j$ 
with timestep $\Delta t_3=\Delta t_0/8$ is $\sim n_0 /10$.
The choice of a simple weighting scheme $w_i \propto N_{OP}(i)$ would 
have given a shape of the load distribution similar to that of $n_j$.
The reason for the shape of the load distribution of Fig. \ref{whist}b is that 
in a
multistep integration scheme the particle forces are calculated within 
a large timestep $\Delta t_0$ when their positions must be synchronized.
An optimal choice of the constraints on the particle timesteps 
yields a binned distribution $n_j$ with a hierarchy $n_{j+1}\sim n_j/2$.
A weighting scheme which sums the number of floating point operations
over a large timestep $\Delta t_0$ takes into account the fact there are
few particles with $\Delta t_j \ll \Delta t_0$ but that these 
particles have forces calculated a number of times $\propto \Delta t_j ^{-1}$.
This weighting scheme leads, at the end of a large timestep $\Delta t_0$, 
to particle loads with a distribution which can be considered roughly constant, 
for practical purposes, for a large fraction of the simulation 
particles ($\sim 90\%$). An ORB domain decomposition is then 
applied every large timestep $\Delta t_0$  according to the 
calculated weights of the particles.
The subdivision of the computational load that follows from this ORB
among the processors is still unbalanced, but within a large timestep 
$\Delta t_0$ the unbalancing is higher when the computational work is
minimal. 

Finally, the load balancing efficiency of the proposed weighting scheme 
was tested against the 
processor number $P$ by measuring  $<L>_n$ for a set of runs.
The binned distribution $n_j$ of Fig. \ref{whist} reveals that there are 
 few particles with timesteps $\Delta t_j = \Delta t_5$, the parallel runs
were performed setting $\Delta t_{min}=\Delta t_0/32$.
The initial conditions for the runs with $P\leq 128$ processors 
were obtained from simulation S6C as previously discussed.
For $P=4$  the relative weights of the different computational works
within a large timestep $\Delta t_0$ are clearly illustrated in 
Fig. \ref{loadh}.
Panel (a) shows the elapsed CPU wall-clock
time spent by the parallel code to compute the gravitational 
forces.  The CPU time is plotted 
 between $t_n$ and $t_{n+1}$ versus the simulation time 
$t_n^{(k)}$ and is the maximum of  the single processor values.
There is a  large burst of CPU work when the particles synchronized 
at $t_n^{(k)}$  are those with timesteps $\Delta t_0$, $\Delta t_0/2$ and
$\Delta t_0/4$.
The instantaneous load balancing $L_{(k)}$ is shown in 
panel (b) and is calculated using Eq. \ref{lbal} 
between $t_n^{(k)}$  and $t_n^{(k+1)}$. 
The CPU times $t_p$ refer to the times spent in the gravity force
computation but without including those necessary for the construction
of the local trees.
It can be seen that $L_{(k)}$
  drops to very inefficient values ($\simlt 0.3$) when
 $t_n^{(k)}$ corresponds to a small number of active particles and it
reaches a high efficiency ($\simgt 0.9$) with the highest CPU times.
The overall load balancing is measured by applying Eq. \ref{lbal} 
between every  ORB domain decomposition; panel (c) shows $<L>_n$ versus 
the simulation time $t_n$ for ten large timesteps $\Delta t_0$.
The ORB procedure was performed setting for the first step $w_i=const$,
 yielding $ <L>\sim 0.8$.
This proves that the load balancing performances are sensitive to the 
chosen weighting scheme and that the procedure previously described is
optimal to achieve a good load balance for the parallel treecode 
described here. 

\begin{figure}[h]
\vspace{-3cm}
\centerline{\mbox{\epsfysize=11.0truecm\epsffile{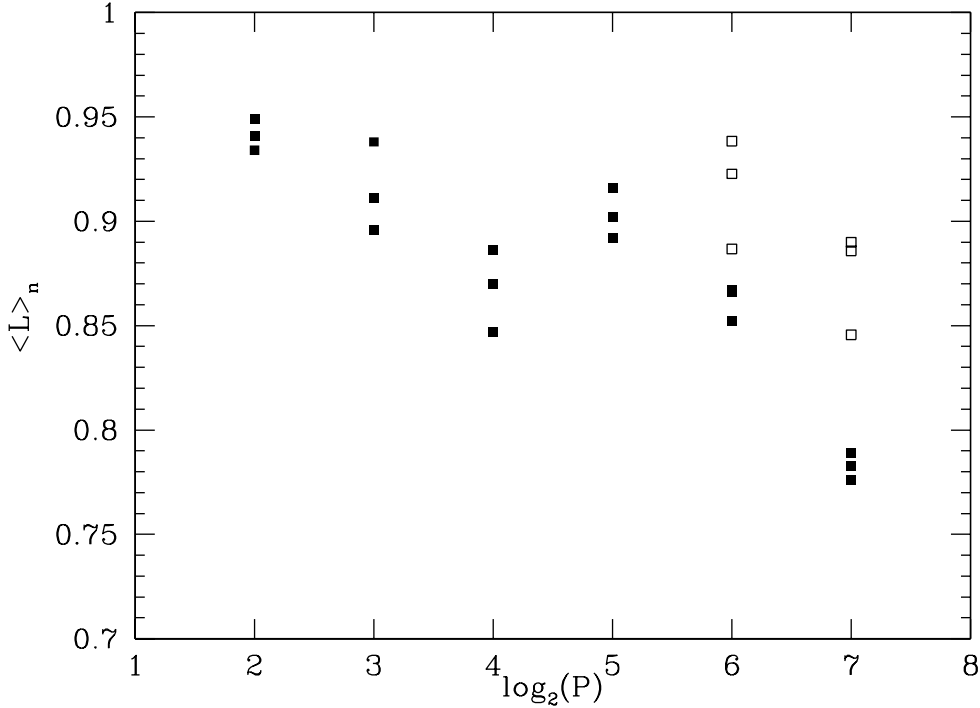}}}
\caption{ 
For a set of parallel runs the values of the load balancing $<L>_n$ at the 
end of a timestep $\Delta t_0$ are shown for three timesteps 
versus the processor number $P$. Initial conditions were obtained from
the serial run $S6C$ (Fig. \ref{whist}), using  the splitting procedure 
described in sect. 3.1.
For $P\geq64$ open symbols refer to the values of $<L>_n$ evaluated  
according to the computational loads used to perform the ORB 
subdivision.
}
 \label{loadb}
 \end{figure}

\begin{figure}
\vspace{-3cm}
\centerline{\mbox{\epsfysize=11.0truecm\epsffile{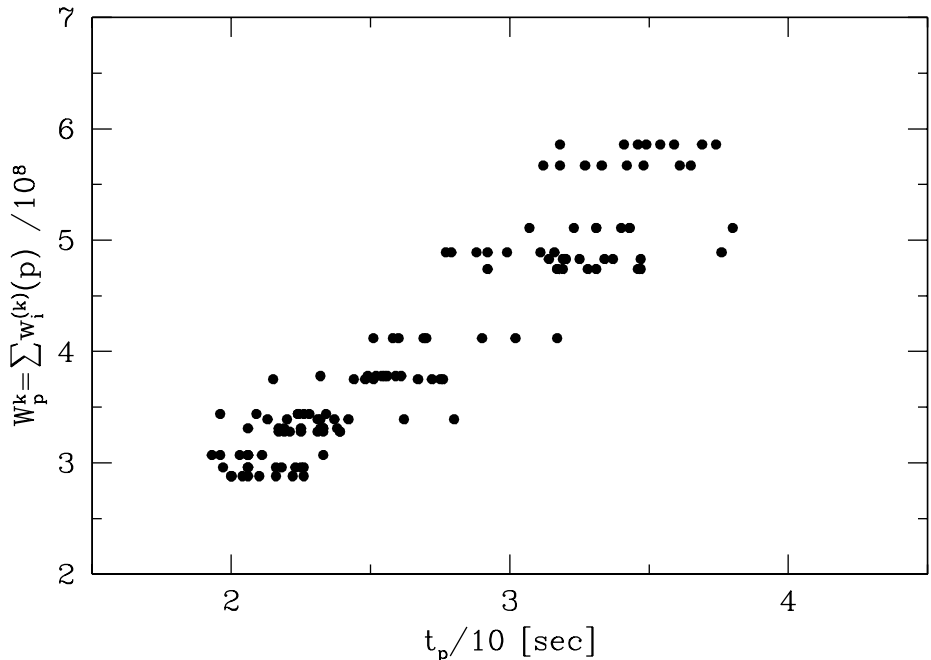}}}
\vspace{-1cm}
\caption{ 
For the parallel run of Fig. \ref{loadb} 
with $P=128$  the computational loads $W_p^k$ of the processors are 
plotted at the simulation time 
$t_n^{(k)}$ versus the corresponding CPU times $t_p$, here $p$ is the 
processor index.  $W_p^k$ is defined as the sum  over the processor
particles of the loads $w_i^{(k)}$. The particle computational loads
$W_i =\sum _k w_i^{(k)}$ 
are those used to perform the ORB procedure. 
The simulation time is chosen at $t_n^{(k)}= 228 \Delta t_0 + \Delta t_0/2$,
when there is the highest number of active particles.
}
 \label{loadw}
 \end{figure}

In Fig. \ref{loadb} the values of $<L>_n$  for the parallel runs are displayed 
versus the 
processor number $P$. For
each value of $P$, the values of $<L>_n$ were computed for three
timesteps $\Delta t_0$. 
The CPU times used for computing $<L>_n$ include also the tree 
construction times.
The measured values of $<L>_n$ demonstrate that this weighting scheme
for the ORB domain decomposition can be successfully used to obtain 
an efficient load balancing, even  when the number of 
processors is high.
These performances are not affected if the measured CPU times used 
for computing $<L>_n$ do not include the tree construction parallel
overheads. For P=32 it is found 
 \tr$(\Delta t_0)/$\tc$(\Delta t_0)\sim 5 \%$ and for $P=128$ this ratio 
grows up to $\sim 10 \%$.
This is an important feature of the parallel treecode presented here,
since in a multistep treecode the tree reconstruction at each timestep can
degrade code performance.
Parallel runs with $P=128$   have 
load balancing down to $L\sim 0.8$.
It must be stressed that, for  $P>32$, these relatively poor values
of the load balancing are not due to a failure of the adopted 
weighting procedure. About a $\sim 10 \%$ of the loss of balancing 
efficiency originates from the scatter of the CPU times 
among different processors with the same amount of computational load.
This is clearly illustrated  by the open symbols of Fig. \ref{loadb}.
For $P\geq 64$ these symbols refer to the values of $<L>_n$ evaluated  
at the end of the step $t_n=n \Delta t_0$ 
using in Eq. \ref{lbal} 
the computational loads 
of each processor $p$ : $W_p\equiv \sum _i W_i$, instead of the corresponding
 CPU times. The particle computational loads
$W_i$ are those used to perform the ORB procedure, hence the load balancing 
measured according to the processor values $W_p$ yields a sort of 
`theoretical maximum'  balancing efficiency for the chosen weighting scheme.
Fig. \ref{loadb} shows that for $P\geq 64$ the values of $<L>_n$ measured 
using the 
computational loads are systematically higher ($\sim 0.9$) than those 
obtained from the CPU times of the processors.
For $P=128$, the processor computational loads versus the corresponding CPU 
times $t_p$ are presented in Fig. \ref{loadw} at the simulation time 
$t_n^{(k)}=228 \Delta t_0 + \Delta t_0/2$. 
The large scatter among the values of $t_p$ for a given computational load 
 shows that the relatively poor code performances in the large $P$ regime are 
not then intrinsic to the proposed weighting method for performing the ORB 
partition, but arise because of the scattering of the CPU times between 
processors 
with the same computational load, which increases as $P$ increases.
The scatter of the CPU times is attributable to the probability of having
the memory access of a process in conflict with concurrent tasks. This 
probability  grows
statistically as $P$ increases and it depends on the machine configuration.

The above runs  have illustrated the load balancing efficiency $L$
 ($\sim 90\%$) for a set of simulations with an evolved clustering state, but 
at high redshifts the values of $L$ for a cosmological simulation are equally 
high. 
This follows because in a multistep integration scheme the particles 
have initially assigned the smallest timestep $\Delta t_{min}$.
The main criterion which determines particle timesteps at early redshifts
is given by Eq. \ref{dt1}. There is thus a large fraction of particles which, 
during the early phases of the simulation, evolves with the same timestep.
For a `monochromatic' distribution function of particle timesteps, 
the load balancing that follows from the ORB procedure is similar to the one
of the single-step integration scheme and is as
high as that of a clustered distribution.
This confirms that the load balancing efficiency of the
adopted weighting scheme is robust for a variety of 
clustering states in cosmological simulations.

 The high load balancing efficiency of the code  compares well 
with the values reported for the gravity computation by 
Dav\'e, Dubinski \& Hernquist (1997) in several tests on the performances of 
their parallel treecode which incorporate SPH hydrodynamics.
For a number of processors $P\leq16$ the obtained values of $L$ are 
higher than  those presented by Springel, Yoshida \&  White (2001, Table 1), 
and similar to the ones of Miocchi \& Capuzzo-Dolcetta (2002, Fig.6).
A proper comparison is difficult however, because the tests of these authors
were performed keeping the number of particles constant, while increasing 
the processor number.
The implementation of SPH hydrodynamics in the parallel treecode is a work 
in progress, however it is unlikely that the  performances of 
the parallel treecode combined with SPH will be degraded.
According to the tests performed by Dav\'e, Dubinski \& Hernquist 
(1997, Table 1),  for $P=8$ processors the SPH computational part of the 
 code is  load balanced  to $\sim97\%$. For the gravity computation, the value 
is similar and in the same range as those found here with the same number of 
processors.

\section{Summary and conclusions}
A parallel treecode which is designed for running cosmological $N-$body 
simulations on parallel machines with distributed memory has been discussed.
The code uses individual timesteps and the gravitational field is solved
according to the BH algorithm.
In this method the accuracy of the individual particle accelerations
is controlled by the value of the opening angle parameter $\theta$.
The code uses a value of $\theta=\theta(t)$ 
which varies with the simulation time $t$ during the integration. 
The optimal value is chosen
to minimize the computational work for evaluating the particle accelerations
while keeping the relative errors below a fixed threshold.
The domain decomposition of the computational volume among the processors 
is performed according to the ORB procedure using a suitable chosen
weighting scheme to measure the particle computational loads.
To compute the gravitational accelerations of the local particles of each 
processor it is necessary to construct first the local essential trees.
The construction of the local essential trees follows that already 
implemented in previous works \cite{du96,da97}. In sect. 2.3 it was seen
that it is possible to concentrate all the communications between processors,
necessary for the construction of the local essential trees, in a single
phase, which can be efficiently managed with a single message-passing
all-to-all routine.

The parallel performances of the code were tested 
on two different machines (IBM SP3 and SP4) and for a set of 
parallel runs up to $P=128$ processors.
 For a fixed value of $\theta$ and keeping constant the number of 
particles per processor, it is found that the CPU time \tc required
to compute the gravitational accelerations is well approximated by 
the expected theoretical scaling relations $\propto log(N_p)$, with a
$\sim  10 \%$ loss in the large $P (\geq 64)$ limit.
The results show that the computational cost \tr necessary 
for the construction of the local trees is a small fraction of the
time \tc. This is important because according to the parallel scheme
presented here the trees must be reconstructed in the processors domains
at each timestep. This could degrade code performance in a multistep scheme
where the number of active particles and then the gravity computational work 
of the code strongly depend on the current timestep.
An important issue related to this point is the load balancing efficiency
of the code when individual timesteps are used. In a multistep integration
scheme, the work load imbalance among the 
processors can be significant at each timestep.
The measured performances show that this task can be efficiently solved 
if the domain decomposition of the computational volume among the processors 
is done using an appropriate weighting of the computational work carried out 
on each 
particle within a large timestep. The results of the 
parallel runs show that the load balancing is  very weakly dependent 
on the particle clustering distribution, and is as high as 
 $\sim 90\%$ up to $P=32$. When a larger number of processors is used, code 
performances are degraded down to $ \sim 80\%$. This relatively low 
efficiency in the large $P$ limit follows because, for a given computational 
load, the scatter
 between the processor CPU times increases as $P$ increases.
If the load balancing is measured according to the computational load of the 
processor particles, it is found that $L \sim 90\%$ up to $P=128$.
This shows that the intrinsic efficiency of the proposed load 
balancing algorithm is as high as $\sim 90\%$ and this is the 
expected value that can be obtained in the large $P$ regime on  a memory 
dedicated machine.

To summarize, it has been shown that 
a multistep parallel treecode can be used to run cosmological simulations 
on massive parallel machines with low communication overheads ($ \simlt 5 \%$), 
a speedup $\sim 10\%$ lower than its theoretical value and a good
load balancing ($\sim 90\%$ up to $P=32$).
The performances of the proposed algorithms for the parallelization of
a treecode should then be considered particularly promising.
This is because the 
continuous growth of computer technology is likely to make 
cosmological simulations on parallel machines with a very large array of 
processors routinely available in forthcoming years.

\section{Acknowledgements}
The author is grateful to C. Cavazzoni, of the CINECA staff, 
 for helpful comments and discussions on parallel algorithms and 
the implementation of the code.

\appendix
\renewcommand{\theequation}{A-\arabic{equation}}
\setcounter{equation}{0}  
 
\section*{Appendix A:Time integration in comoving coordinates}

The equation of motions of the particles are defined in comoving coordinates 
as follows:

\be
\begin{array}{lll}
\frac{d \vec x}{ d t}&=&\vec v , \\
\frac{d \vec v}{d t}&=&\frac {\vec g }{a(t)^3}-2 H(t) \vec v ,
\end{array}
\ee
where $\vec r= a(t) \vec x$ is the proper distance, $a(t) \vec v$ is the
peculiar velocity and  $\vec g /a^2$ is the peculiar 
acceleration.  It is useful to introduce the notation  
$\vec g \prime \equiv \vec g /a^3$.  
The particle positions and velocities are advanced according to a second
order leapfrog integrator. The spatial coordinates at the step $n+1/2$ and
the velocities at the step $n+1$ are then, respectively, given by 

\be
\begin{array}{lll}
\vec x_{n+1/2}&=&\vec x_{n-1/2} +\Delta t\vec v_n, \nonumber \\
\vec v_{n+1}~~&=&\vec v_{n}+\Delta t \left [\vec g \prime -2 H \vec v \right]_{n+1/2}
 = \\
~~~~~~~~~~~~~~&=&\vec v_{n}\left[ \frac{1-\Delta t H}
{1+\Delta t H} \right]_{n+1/2}
+\Delta t \left [ \frac{\vec g \prime}{1+\Delta t H}\right]_{n+1/2}, 
\end{array}
\ee
where the variables in square brackets are evaluated at the timestep indicated
 by the 
subscript and $\vec v_{n+1/2}= (\vec v_{n+1}-\vec v_n)/2$. 
At the beginning of the integration all of the particles 
have the same timestep $\Delta t= \Delta t_{min}$, and
 the particle positions  are centered at half time step according to
\be
\vec x_{n+1/2}=\vec x_{n} +\frac {\Delta t}{2}\vec v_n+
\frac{\Delta t^2} {8}\left [\vec g \prime -2 H \vec v \right]_{n}.
\ee
If it is necessary to synchronize at the step $n+1$ the particle positions
with the velocities , then $\vec x_{n+1}$ is given by

\be
\begin{array}{lll}
\vec x_{n+1}&=&\vec x_{n+1/2} +\frac {\Delta t}{2} \vec v_{n+1/2}+ 
\frac{\Delta t^2} {8}\left [\vec g \prime -2 H \vec v \right]_{n+1/2}=\\
~~~~~~~~~~~~&=&\vec x_{n+1/2}+\frac {\Delta t}{2} \vec v_{n+1}+
\frac {\Delta t^2}{4}\vec v_{n+1} 
\left [\frac {H}{1-H \Delta t}\right]_{n+1/2}\\
~~~~~~~~~~~~&-&\frac{\Delta t^2} {8}
\left [\vec g \prime \frac{1+ H \Delta t}{1-H \Delta t}\right ]_{n+1/2},
\end{array}
\ee
where the mid-point velocity $\vec v_{n+1/2}$ is estimated using 

\begin{eqnarray}
~~~~~\vec v_{n+1}=\vec v_{n+1/2} +\frac {\Delta t}{2}  
\left [\vec g \prime -2 H \vec v \right]_{n+1/2}.  
\end{eqnarray}

When a particle changes its timestep from $\Delta t_{old}$ to $\Delta t_{new}$,
 its position must be corrected 
in order to preserve second order accuracy. The correction terms are 

\be
\vec x \rightarrow \vec x + \left (\frac{\Delta t_{new}^2}{8}  
- \frac{\Delta t_{old}^2}{8} \right )  
\left [\vec g \prime -2 H \vec v \right]_{n+1/2},  
\ee
the subscript of the term in square brackets refers to the last time 
the acceleration of the particle has been evaluated. Using Eq. A.5 this term
can be written as 

\be
\left [\vec g \prime -2 H \vec v \right]_{n+1/2} =
\left [\frac {\vec g \prime} {1-H \Delta t} \right]_{n+1/2}-
\left [\frac {2H} {1-H \Delta t} \right ]_{n+1/2} \vec v_{n+1}.
\ee

\end{document}